\documentclass[a4paper, twoside]{article}
\usepackage{amsmath}
\usepackage{amssymb}

\def\({\left(}
\def\){\right)}
\def\be{\begin{equation}}
\def\ee{\end{equation}}
\def\b#1{\mathbf{#1}}
\def\bx{\b x}
\def\br{\b r}
\def\bX{\b X}
\def\l{{(\lambda)}}
\def\tilde{\widetilde}
\def\Hat#1{\ifcat 0#1\hat{\vphantom{k}#1}\else{\kern1.5pt\hat{\vphantom{k}\kern-1.5pt#1\kern0pt}}\fi}
\def\d{{\rm d}}

\begin{document}

\thispagestyle{empty}
\markboth{On The Equivalence of the FRW Field Equations and those of Newtonian Cosmology\hfil \hfil\hfil}{\hfil C.G.~Wells}
\pagestyle{myheadings}

\arraycolsep=2pt

\centerline{\huge On The Equivalence of the FRW Field}
\vskip5pt
\centerline{\huge Equations and those of Newtonian Cosmology}
\vskip1cm
 
\centerline{\sc Clive~G.~Wells}
\vskip5pt
\centerline{\it\small Centre for Mathematical Sciences, Wilberforce Road, Cambridge, CB3 0WA, United Kingdom.}

\centerline{{\it\small Email address:\/} {\tt C.G.Wells{\rm@}damtp.cam.ac.uk}}

\vskip5mm

\setcounter{footnote}{0}

\vskip0.3cm

\vskip1cm

\begin{abstract}
We present a simple argument to explain why the field equations of the Friedmann-Robertson-Walker metric are equivalent to those of Newtonian cosmology. By passing to the infinite limit of a family of conformally rescaled FRW metrics in suitable coordinates, we reveal Newtonian space and time. The limiting process preserves the Einstein equations and these may be elucidated directly from the Newtonian limit up to the determination of the scalar curvature parameter. Consideration of the conformally invariant scalar field equation on the FRW spacetime is used to  recover the Einstein equations efficiently from the Newtonian theory.  We proceed to examine the limiting procedure in connection with the Cartan formulation of Newtonian gravity.


\end{abstract}

\section{Introduction}

Since the work of Milne~\cite{Milne} and  McCrea \& Milne~\cite{McCreaMilne} in 1934 it has been known that the equations governing the evolution of the FRW metric in the general theory of relativity and those governing the evolution of Newtonian cosmology are formally identical except for the interpretation of a single parameter. This piece of apparent good fortune has allowed introductory courses on cosmology to be taught at undergraduate level without the prerequisite requirement of the student undertaking a course of study in general relativity.

In this paper we give a straightforward conformal rescaling argument that explains why the governing equations of the relativistic and Newtonian theories are identical. The idea is to rescale the curvature parameter and the speed of light in such a way that the Einstein equations are preserved. In the limit that the scaling becomes arbitrarily large Newtonian space and time are made manifest. The argument used can be understood in terms of  a conformal rescaling of the metric. Another way to think about the process is as blowing up a small neighbourhood of an arbitrary point, so that the scalar curvature becomes negligible and the speed of light tends to infinity, in this way the analysis is related to Manton's work on constructing a Newtonian atlas for the FRW metric~\cite{Manton}. In that paper, Manton expressed the FRW metric in terms of geodesic coordinates to reveal the local Newtonian spacetime, in the context of the weak field limit of general relativity, up  to quadratic order.

The plan of the paper is as follows. In section~\ref{sect:conformal} we present a family of conformally related metrics and introduce a set of coordinates that lead to Newtonian space and time in the limit that the conformal parameter tends to infinity. We identify certain geometrical objects that have well-defined Newtonian limits and establish what information is lost when we pass to the limit. In section~\ref{sect:Newton} we give a reprise of a conventional Newtonian argument often used as a derivation of the Friedmann and Raychaudhuri equations. The argument is simple but flawed and has a long history of criticism in the literature. However the equations derived are correct, if for more sophisticated reasons than presented. In section~\ref{sect:k} we see how to use the Newtonian equations to derive the Einstein equations for the FRW metric. This involves determining the value of the one piece of information lost when the Newtonian limit is taken. This piece of information is the value of the scalar curvature parameter. We do this by considering a solution of the conformally invariant scalar field equation on the FRW spacetime. Using the observation that the Weyl tensor vanishes and so the FRW metric is (locally) conformally flat we deduce that the Newtonian potential of a point mass solves this equation away from the point mass itself and the Ricci scalar of the metric may be computed with minimum effort. This links the one lost piece of information from the Newtonian limit with its analogue in the Newtonian theory.

In section~\ref{sect:Cartan} we relate the limiting process to Cartan's formulation of Newtonian gravity. Many of the ingredients of that theory arise simply as the limit of the relativistic counterparts of our conformally related metrics. The use of Cartan's theory of Newtonian gravity removes the difficulties associated with the more elementary approaches referred to section~\ref{sect:Newton}. In the appendix to the paper we show how the Newtonian results and the deduction from section~\ref{sect:k} allow us to write down the Riemann tensor of the FRW metric with almost no computation. The derivation also shows why the Weyl tensor vanishes, a result used in (but not depending on) section~\ref{sect:k}.

\section{The Conformal Rescaling Argument}
\label{sect:conformal}

The FRW metric may be expressed in isotropic coordinates by the form~\cite{Bondi}
\be
\d s^2=-c^2\d t^2+\frac{a(t)^2\d\bX^2}{(1+\frac14k\bX^2)^2}.\label{eq:1}
\ee
We will consider this metric on the chart $(t,\bX)\in\mathbb{R}\times\mathbb{R}^3$. Note that for $k\neq0$ the metric appropriately defined on the one point compactification of the space  possesses an isometry, which is inversion through the sphere of radius $|\bX|=2|k|^{-\frac12}$.
For $k<0$ the metric is singular on this sphere and ordinarily this defines the natural boundary of the spacetime (the spatial metric is the hyperbolic metric on a ball). In this paper it will be convenient to allow a singular metric to be defined on this chart.\footnote{There is some merit in recasting the argument that follows in terms of the inverse metric $g^{ab}$ which  then has finite components at all points. However doing so obscures some of the more physical aspects of the limiting process.}

The FRW metric given by Eq.~(\ref{eq:1}) can be embedded into a one-parameter family of metrics given by
\be
\d s_\l^2=-c^2\lambda^2\d t^2+\frac{a(t)^2\d\bx^2}{(1+\frac14k\lambda^{-2}\bx^2)^2}.\label{eq:2}
\ee
The change of coordinates
$\bx=\lambda\bX$
brings the metric into the form
\be
\d s_\l^2=\lambda^2\(-c^2\d t^2+\frac{a(t)^2\d\bX^2}{(1+\frac14k\bX^2)^2}\)\label{eq:3}
\ee
and so our family is a collection of conformally rescaled metrics. The Ricci tensors\footnote{
The conventions for the Riemann and Ricci tensors used in this paper are: $R^a{}_{bcd}Z^b=(\nabla_c\nabla_d-\nabla_d\nabla_c)Z^a$ and $R_{ab}=R^c{}_{acb}$.} of the conformal metrics are related by
$R^\l{}^a{}_b=\lambda^{-2}R^a{}_b$.
Now let us impose the Einstein equations which only need to hold in a fixed neighbourhood of $\bX=\b0$:
\be
R^\l{}^a{}_b=\lambda^{-2}R^a{}_b=\frac{8\pi G}{c^4\lambda^2}\(T^a{}_b-{
\textstyle\frac12}T^c{}_c\,\delta^a{}_b\).\label{eq:4}
\ee
For a perfect isentropic fluid the energy-momentum tensor is given by\footnote{We may represent the presence of a cosmological constant by using a contribution to the energy and pressure due to dark energy.} $(T^a{}_b)={\mbox{\rm{diag}}}(-\rho c^2,p,p,p)$.
Since $R^{\l}_{ab}=R_{ab}$, the Einstein equations determining $a(t)$ are independent of $\lambda$, and accordingly  $a(t)$ may be taken to be independent of $\lambda$. This is fortunate, since our conformal rescaling would have failed if the imposition of the Einstein equations had required that $a(t)$ depend on $\lambda$.

We will work now in the $(t,\bx)$ coordinate system. Lowering the first index,\footnote{We use indices from the middle part of the alphabet to represent the spatial coordinates.}  of Eq.~(\ref{eq:4}) 
gives
\be
R^\l_{tt}=4\pi G\(\rho+\frac{3p}{c^2}\);\quad R^\l_{ti}=0\quad\mbox{and}\quad R^\l_{ij}=\frac{4\pi G(\rho c^2-p)a^2\delta_{ij}}{c^4\lambda^{2}(1+\frac14k\lambda^{-2}\bx^2)^2}.\label{eq:5}
\ee

Next examine the properties of the metric given by Eq.~(\ref{eq:2}). The spatial part of the metric,
\be
\d r_\l^2\equiv\frac{a(t)^2\d\bx^2}{(1+\frac14k\lambda^{-2}\bx^2)^2}\to a(t)^2\d\bx^2\qquad\mbox{as $\lambda\to\infty$,}
\ee
the limit being taken at constant $\bx$. This is a time-dependent flat space. Observe that the speed of light  measured with respect to the $t$-coordinate is
\be
\frac{\d r_\l}{\d t}=c\lambda\to\infty
\ee
as $\lambda\to\infty$. This is our Newtonian limit.

Let $V=\partial/\partial t$ be the tangent vector of a congruence of timelike geodesics in the spacetime with metric given by Eq.~(\ref{eq:2}), the equation of geodesic deviation reads
\be
\frac{\d^2Z^a}{\d t^2}+E^{\l a}{}_bZ^b=0,\qquad
E^{\l a}{}_b=-R^{\l a}{}_{cdb}V^cV^d,
\ee
where $E^{\l a}{}_b$ is the tidal tensor, with spatial components, $E^{\l i}{}_j=\frac13R_{tt}\delta^i{}_j$ by isotropy. We will need this observation later to restrict the possible Newtonian potentials arising in the Newton-Cartan theory. It is relevant to point out that the tidal tensor has a well-defined analogue in the Newtonian theory.

Although the Einstein equations hold for all finite values of $\lambda$, 	and moreover are independent of $\lambda$, not all of the equations can be recovered directly from the Newtonian limit. Only those for which there is an analogous Newtonian counterpart are available to us. In particular the Einstein equations governing $R^\l_{ij}$ do not have a well-defined Newtonian analogue. This is to be expected since this equation depends on the value of $k$, the scalar curvature parameter, and there is no counterpart in the Newtonian theory; indeed, one finds the same Newtonian limit for all values of $k$. However note that the contracted Bianchi identity implies that
\be
-\frac1{a^6c^2}\frac{\d}{\d t}\(a^6R^{\l}_{tt}\)=\frac{\lambda^2}{a^6}\frac{\d}{\d t}\(a^6R^{\l t}{}_t\)=\frac{\lambda^2}{a^2}\frac{\d}{\d t}\(a^2R^{\l i}{}_i\),
\ee
which is equivalent to the conservation of the energy-momentum tensor, i.e., \mbox{$\nabla_bT^{ab}=0$} and hence much of the information about the spatial components of the Ricci tensor are encoded into the fluid conservation equation and the tidal tensor both of which do have Newtonian counterparts.

It is worth elaborating how the conformal rescaling works. The underlying manifold has been given a chart defining the $(t,\bX)$ coordinates. On this chart there is a family of conformal metrics indexed by the parameter $\lambda$. The Einstein equations are invariant under this conformal change and it is important to point out that these equations are independent of position due to the spatial homogeneity of the FRW metric. When we change coordinates to $(t,\bx)$ we make a $\lambda$-dependent change and the metrics are no longer conformal (since we would be identifying different points in the original $(t,\bX)$ coordinate system). However, since the FRW metric is spatially homogeneous the change of coordinates does not change the Einstein equations and it is the interplay of the two coordinate systems that allows us to deduce the final result

\section{The Conventional Newtonian Argument}
\label{sect:Newton}

In this section we reproduce a simple argument, originally due to Milne~\cite{Milne}, presented in many standard cosmology textbooks;  we will consider a more sophisticated approach in section~\ref{sect:Cartan}. The proof below makes use of the Newtonian result applicable for spherically symmetric {\em finite\/} mass distributions that the gravitational attraction experienced by a particle is determined entirely by the mass within a spherically symmetric ball on which the test particle resides. A  better approach, as expounded by McCrea~\cite{McCrea}, is to construct a Newtonian potential solving the Poisson equation. However the boundary conditions appropriate for an infinite mass distribution are not altogether clear, see Layzer~\cite{Layzer2} and the response due to Heckmann \& Sch\"ucking~\cite{HeckmannSchucking}.\footnote{Heckmann \& Sch\"ucking postulate  boundary conditions on the Newtonian tidal tensor, $E_{ij}=\nabla_i\nabla_j\phi$ given by \mbox{$E^i{}_j-\frac13\delta^i{}_jE^k{}_k=A^i{}_j(t)$} as $|\br|\to\infty$. where $A^k{}_k(t)=0$.
These boundary conditions allow for anisotropic Newtonian solutions (e.g., Newtonian analogues of the Kantowski-Sachs solution~\cite{KantowskiSachs}). Although interesting, we remark that these boundary conditions are consistent with our assumptions of homogeneity and isotropy of the cosmological solution only if $A^i{}_j(t)=0$.}

Consider a comoving test particle of mass $m$ located at a position $\br=a(t)\bx$ relative to a comoving observer  defining the origin. Next we examine the gravitational attraction experienced by the test particle due to the ball of fluid of radius $|\br|$ centred at the observer. Conservation of energy\footnote{Conservation of energy is seen to hold since we can construct a time-independent Lagrangian given by \mbox{$L=\frac12 m\(\dot a^2+\frac83\pi G\rho a^2\)\bx^2$}, where $\rho$ should be regarded as a function of $a$ obeying $\d\rho/\d a=-3(\rho+p/c^2)/a$, i.e., Eq.~(\ref{eq:10}).} requires that
\be
\frac12 m\dot\br^2-\frac{GMm}{|\br|}=\frac12 m\(\dot a^2-\frac83\pi G\rho a^2\)\bx^2=-\frac12 m\tilde kc^2\bx^2
\ee
since $M=\frac43\pi\rho|\br|^3$, the gravitational attraction being determined only by the mass within the 
ball,  here $\tilde k$ is a constant proportional to the conserved energy, for consistency it is independent of $\bx$. Rearranging this equation leads to the Friedmann equation:
\be
\dot a^2+\tilde kc^2=\frac{8\pi G\rho a^2}3.\label{eq:9}
\ee
To proceed further we use the First Law of Thermodynamics. Let us point out that the first law is implicit in the construction of the energy-momentum tensor for a perfect isentropic fluid, see Hawking \& Ellis~\cite{HawkingEllis}, pp.\,69-70 for the construction. We have, from special relativity\footnote{Here $c$ should be regarded as just a constant, since in the Newtonian treatment the speed of light is infinite.} and the first law that, $\d E=\d(\rho c^2V)=-p\,\d V$ and thus
\be
\dot\rho=-\frac{\dot V}V\(\rho+\frac p{c^2}\)=-\frac{3\dot a}a\(\rho+\frac p{c^2}\),\label{eq:10}
\ee
since comoving volumes are proportional to $a^3$. Differentiating Eq.~(\ref{eq:9}) yields the Raychaudhuri equation:
\be
\ddot a=-\frac{4\pi G}3\(\rho+\frac{3p}{c^2}\)a.\label{eq:11}
\ee
Our conformal rescaling argument shows these equations 
are identical with the Einstein equations applied to the FRW metric, Eq.~(\ref{eq:1}), provided one can determine the value of $\tilde k$.

\section{Determination of $\tilde k$}
\label{sect:k}

In the Newtonian treatment the constant of integration $\tilde k$ appears. Clearly this constant is determined by the metric, Eq.~(\ref{eq:1}) in some way. We will find that $\tilde k=k$, the spatial curvature parameter, it is hopeless to try to find this from the Newtonian theory since the limit of all FRW families is independent of $k$. We need to work with the FRW metric itself. To this end consider the problem of solving the conformally invariant scalar field equation, $\square\,\phi-\frac16R\phi=0$ on the FRW spacetime. Since the Weyl tensor vanishes (see the appendix), 
the spacetime is locally conformally flat (pp.\,137-139 of Penrose \& Rindler~\cite{PenroseRindler}), i.e., there exists $\Omega(\bx,\tau)$ defined in a neighbourhood of the origin such that
\be
\d s^2=-c^2\d t^2+a(t)^2\left[\frac{\d r^2}{1-kr^2}+r^2(\d\theta^2+\sin^2\theta\,\d\varphi^2)\right]=\Omega^2\left[-c^2\d\tau^2+\d\bx^2\right].
\ee
The conformally invariant scalar field equation has solution $\phi=\tilde\phi/\Omega$ where $\tilde\phi$ obeys the wave equation $\tilde\square\,\tilde\phi=0$ with respect to the metric $\d \tilde s^2=-c^2\d\tau^2+\d\bx^2$. We take $\tilde\phi=1/|\bx|$ as a solution for $\bx\neq\b 0$ and hence $\phi=1/\Omega|\bx|=1/ar$, i.e., the Newtonian potential of a point mass, solves the conformally invariant scalar field equation away from the origin. Accordingly the Ricci scalar for the FRW metric is given by
\begin{eqnarray}
R=6\phi^{-1}\square\,\phi
&=&-\frac{6}{a^2c^2}\frac\d{\d t}\(a^3\frac\d{\d t}\(\frac1a\)\)
+
\frac{6(1-kr^2)^{\frac12}}{a^2r}\frac\d{\d r}\(r^2(1-kr^2)^{\frac12}\frac{\d}{\d r}\(\frac1r\)\)\\
&=&\frac 6{c^2}\(\frac{\ddot a}a+\frac{\dot a^2}{a^2}+\frac{kc^2}{a^2}\).
\end{eqnarray}
Substitution from the Friedmann and Raychaudhuri equations,  Eqs.~(\ref{eq:9}) \& (\ref{eq:11}) on the one hand and from the trace of the Einstein equation, $R=8\pi G(\rho c^2-3p)/c^4$ on the other, implies that $\tilde k=k$.

\section{Newton-Cartan Theory}
\label{sect:Cartan}

Cartan's formulation of Newtonian gravity has been advocated ~\cite{Malament,Malament2,Malament3,Tipler,Straumann} as a way to resolve the ambiguities of the more elementary treatments of the cosmological aspects of  Newton's theory of gravity. 
In this section we identify the Newton-Cartan structure that arises naturally from the conformal limit we have been examining. Our notation follows that of~\cite{Malament}. Cartan's Newtonian structure is $(t_{ab}, h^{ab},\nabla,\mathop\nabla\limits^\phi,{\cal M})$, where $\cal M$ is a four-dimensional manifold, $t_{ab}$ is a temporal metric, $h^{ab}$ is an inverse spatial metric, $\nabla$ is a flat connection on $\cal M$ and $\mathop\nabla\limits^\phi$ is a connection on $\cal M$ such that the geodesics with respect to this connection determine the Newtonian acceleration. We will take ${\cal M}\simeq \mathbb{E}^4$, the four-dimensional affine space.

{}From the metric given by Eq.~(\ref{eq:2}), we construct $t_{ab}=t_at_b$ with $t_a=\partial_a t$ using the coordinates $(t,\bx)$, and define the spatial metric $(h^{ab})=\lim\limits_{\lambda\to\infty}(g^{\l ab})=\mbox{\rm diag}(0,a^{-2},a^{-2},a^{-2})$. Proceed by defining 
 $\mathop R\limits^\phi{}_{ab}=\lim\limits_{\lambda\to\infty}R^\l_{ab}$ and see that it obeys the tensorial relationship
\be
\mathop R\limits^\phi{}_{ab}=4\pi G\(\rho+\frac {3p}{c^2}\)t_{ab}\label{eq:16}
\ee
by virtue of Eq.~(\ref{eq:5}).  
At this point it is useful to define  $\br=a(t)\bx$ so that with respect to the $(t,\br)$
coordinates, $(h^{ab})={\rm diag}(0,1,1,1)$ and $(t_{ab})={\mbox{diag}}(1,0,0,0)$.

Let the 
Newtonian potential be
\be
\phi(\br,t)=\frac{2\pi G}3\(\rho+\frac{3p}{c^2}\)(\br-\br_0(t))^2+f(t).\label{eq:17}
\ee
This is the general solution of
\be
h^{ab}\nabla_a\nabla_b\phi=4\pi G\(\rho+\frac{3p}{c^2}\)\label{eq:18}
\ee
subject to the same isotropy condition on the Newtonian tidal tensor, $\mathop E\limits^\phi{}^{ab}=h^{ac}h^{bd}\nabla_c\nabla_d\phi\propto h^{ab}$ as its relativistic counterpart obeys.

We are now in a position to define
the new affine connection $\mathop\nabla\limits^\phi$  by
\be
\mathop\nabla\limits^\phi{}_aA^b=\nabla_aA^b+t_{ac}h^{bd}\nabla_d\phi A^c
\ee
where with our choice of coordinates, $\nabla_0=\partial/\partial t$ and $\nabla_i=\partial/\partial r^i$ (observe  that the compatibility and orthogonality conditions: $\nabla_ah^{bc}=\mathop\nabla\limits^\phi{}_ah^{bc}=\nabla_a t_{b}=\mathop\nabla\limits^\phi{}_at_b=0$ and $h^{ab}t_b=0$ are all satisfied). It is an elementary computation to show that the  Ricci tensor derived from $\mathop\nabla\limits^\phi$ is then $\mathop R\limits^\phi{}_{ab}=h^{cd}\nabla_c\nabla_d\phi \,t_{ab}$ consistent with Eqs.~(\ref{eq:16})~\&~(\ref{eq:18}) and the tidal tensor is $\mathop E\limits^\phi{}^{ab}=\mathop R\limits^\phi{}^a{}_{cde}h^{bd}V^cV^e=\lim\limits_{\lambda\to\infty}E^{\l ab}$ with $V=\partial/\partial t$.

In Newton-Cartan theory the acceleration of test particles under the influence of gravity is determined by the geodesic equation $V^b\mathop\nabla\limits^\phi{}_bV^a=0$ using the new affine connection, where $(V^a)=(1,\dot\br(t))$ is the 4-velocity of a particle moving along the geodesic. In particular, for the comoving fluid the geodesic equation leads to
\be
\ddot\br(t)=
-\nabla\phi=-\frac{4\pi G}{3}\(\rho+\frac{3p}{c^2}\)(\br(t)-\br_0(t)).\label{eq:20}
\ee
If we take $\br_0(t)$ to be convected with the cosmological fluid then $\ddot\br_0(t)=\b0$ and  
we write $\br(t)-\br_0(t)=a(t)\bar\bx$, i.e., we consider the expansion to be centred at $\br_0(t)$
and hence
\be
\ddot a\bar\bx=-\frac{4\pi G}{3}\(\rho+\frac{3p}{c^2}\)a\bar\bx.\label{eq:21}
\ee
The LHS represents the acceleration relative to an observer at $\b r_0(t)$ (see Norton~\cite{Norton}), 
We observe that Eq.~(\ref{eq:21}) is the Raychaudhuri equation, i.e., Eq.~(\ref{eq:11}) once again.
By using the fluid conservation equation, Eq.~(\ref{eq:10}), we can integrate Eq.~(\ref{eq:21}) to derive the Friedmann equation with $\tilde kc^2$ being a first integral of the system.

\section{Concluding Remarks}

In this paper we have shown how the equations governing the evolution of the scale factor of Newtonian cosmology arise as the limit of a  conformally related family of FRW spacetimes. The Einstein equations for each member of this family are identical and are preserved in the limit with precisely one exception. One piece of information, the scalar curvature parameter of the FRW metric is lost. In this way the FRW field equations may be extracted from the Newtonian theory up to the identification of the scalar curvature parameter $k$. It is important to emphasize that in contrast to just taking $c\to\infty$ the Einstein equations are preserved and hence the scale factor is unchanged as the limit is taken. Furthermore the limit is Newtonian space and time (as opposed to a curved space). In this way our conformal rescaling argument explains {\em why\/} the equations of Newtonian cosmology and the FRW field equations are equivalent, i.e., the scale factor of the relativistic theory is identical to the scale factor of the Newtonian theory. This conclusion does not follow from taking the limit $c\to\infty$ on its own (unless $k=0$).

We have shown that by considering a solution of the conformally invariant scalar field equation on the FRW spacetime that the constant of integration found in the Newtonian treatment may be related to the quantity $k$ and hence the complete set of Einstein equations for the FRW spacetime may be written down. Furthermore using this additional piece of knowledge the entire Riemann tensor may be found from the Newtonian results (see the appendix).

We have seen that the gravitational theory of the limiting process we have taken is closely related to Cartan's description of Newtonian gravity. This seems to be the most effective way to resolve the ambiguities present in the more elementary treatments of Newton's theory.

Our method sheds some light on the reason why McCrea's~\cite{McCrea} response to Layzer~\cite{Layzer} was correct, at least in certain cases, with his suggestion that Newton's theory for a uniform infinite mass distribution should be interpreted as that  due to the mass within a finite ball in the limit that its radius becomes infinite.  At least when the solution possesses a cosmological horizon, each of the family of FRW metrics has a finite distance over which the mass distribution can influence the evolution of the system. This distance becomes infinite as the Newtonian limit is reached.  When there is no cosmological horizon the argument is somewhat less clear-cut and one might prefer to understand Newtonian theory in relation to the work originated by Cartan~\cite{Cartan,Cartan2} and Friedrichs~\cite{Friedrichs}, with important insights later by Trautman~\cite{Trautman} and Dixon~\cite{Dixon}.

\noindent{\bf Acknowledgement:} I would like to thank Nick Manton for a useful discussion about reference~\cite{Manton} that helped to raise and clarify some of the ambiguities surrounding  Newtonian cosmology.

\section*{Appendix: Components of the Riemann Tensor}
\label{sect:Riemann}

If one is prepared to accept the results of the Newtonian argument in section~\ref{sect:Newton}, 
or use the rigorous exposition in section~\ref{sect:Cartan}
and the determination of $\tilde k$ from  section~\ref{sect:k}, then all the components of the Riemann tensor for the FRW metric are readily computed. Firstly the Einstein equations give, with respect to an orthonormal basis with the zeroth vector orthogonal to the hypersurfaces of constant $t$,
\be
G_{\Hat0\Hat0}=\frac{8\pi G}{c^2}\rho\qquad\mbox{and}\qquad
G_{\Hat i\Hat j}=\frac{8\pi G}{c^4}p\,\delta_{\Hat i\Hat j}.
\ee
Homogeneity and isotropy imply that $G_{\Hat0\Hat i}=0$ and $G_{\Hat i\Hat j}\propto\delta_{\Hat i\Hat j}$ at all points without computation. It follows from the Friedmann and Raychaudhuri equations, Eqs.~(\ref{eq:9}) \& (\ref{eq:11}) respectively, that 
\be
G_{\Hat0\Hat0}=\frac{3\dot a^2}{a^2c^2}+\frac{3k}{a^2}\qquad\mbox{and}\qquad
G_{\Hat i\Hat j}=-\(\frac{2\ddot a}{ac^2}+\frac{\dot a^2}{a^2c^2}+\frac k{a^2}\)\delta_{\Hat i\Hat j},
\ee
and thus
\be
R_{\Hat0\Hat0}=-\frac{3\ddot a}{ac^2}\qquad\mbox{and}\qquad
R_{\Hat i\Hat j}=\(\frac{\ddot a}{ac^2}+\frac{2\dot a^2}{a^2c^2}+\frac {2k}{a^2}\)\delta_{\Hat i\Hat j}.
\ee
Homogeneity and isotropy also allow us to write down the components of the Riemann tensor at any point, the well-known isotropic forms for cartesian tensors in three dimensions lead to
\be
R_{\Hat0\Hat i\kern0.5pt\Hat0\Hat j}=A\delta_{\Hat i\Hat j};\qquad
R_{\Hat0\Hat i\Hat j\Hat k}=B\epsilon_{\Hat i\Hat j\Hat k}\qquad\mbox{and}\qquad
R_{\Hat i\Hat j\Hat k\Hat l}=C(\delta_{\Hat i\Hat k}\delta_{\Hat j\Hat l}-\delta_{\Hat i\Hat l}\delta_{\Hat j\Hat k}),
\ee
for scalars $A$, $B$ and $C$. We have used the symmetries of the Riemann tensor on the final expression to replace the three undetermined scalar quantities by one.

Now $R_{\Hat0\Hat i\kern0.5pt\Hat0\Hat i}=R_{\Hat0\Hat0}$ implies that $A=\frac13
R_{\Hat0\Hat0}$, $R_{\Hat0[\Hat i\Hat j\Hat k]}=0$ implies that $B=0$ and $R_{\Hat i\Hat j\Hat i\Hat j}=R+2R_{\Hat0\Hat0}$ implies that $C=\frac16(R+2R_{\Hat0\Hat0})$, i.e.,
\be
R_{\Hat0\Hat i\kern0.5pt\Hat0\Hat j}=-\frac{\ddot a}{ac^2}\delta_{\Hat i\Hat j};\quad
R_{\Hat0\Hat i\Hat j\Hat k}=0\quad\mbox{and}\quad
R_{\Hat i\Hat j\Hat k\Hat l}=\(\frac{\dot a^2}{a^2c^2}+\frac k{a^2}\)(\delta_{\Hat i\Hat k}\delta_{\Hat j\Hat l}-\delta_{\Hat i\Hat l}\delta_{\Hat j\Hat k}).
\ee
Observe that the same argument, with the Ricci tensor replaced by zero, allows us to see immediately that the Weyl tensor, $C_{abcd}=0$. The vanishing of the Weyl tensor may be understood as a direct consequence of the fact the FRW metric is conformally flat~\cite{Iihoshi,Ibison}.


\begin{thebibliography}{9}


\bibitem{Milne} E.A.~Milne, Quart.~J.~Math.~{\bf5}:64-72  (1934)

\bibitem{McCreaMilne} W.H.~McCrea \& E.A.~Milne, Quart.~J.~Math.~{\bf5}:73-80  (1934)

\bibitem{Manton} N.S.~Manton, DAMTP preprint (DAMTP-2013-70), {[gr-qc/1312.6040]} (2013)


\bibitem{Bondi} H.~Bondi, {\em Cosmology,\/} Second Edition, Cambridge University Press, Cambridge (1960) 

\bibitem{McCrea} W.H.~McCrea, Ast.~J.~{\bf60}:271-274 (1955)

\bibitem{Layzer2} D.~Layzer, The Observatory {\bf76}:73-74 (1956)

\bibitem{HeckmannSchucking} O.~Heckmann \& E.~Sch\"ucking, The Observatory {\bf76}:74-75 (1956)

\bibitem{KantowskiSachs} R.~Kantowski \& R.K.~Sachs, J.~Math.~Phys.~{\bf7}(3):443-446 (1966)

\bibitem{HawkingEllis}  S.W.~Hawking \& G.F.R.~Ellis, {\em The Large Scale Structure of Spacetime,\/} Cambridge University Press, Cambridge (1973) 

\bibitem{PenroseRindler} R.~Penrose \& W.~Rindler, {\em Spinors and Space-Time: Volume 2, Spinor and Twistor Methods in Space-Time Geometry,\/} Cambridge University Press, Cambridge (1986) 


\bibitem{Malament} D.~Malament, Philosophy of Science {\bf 62}(4):489-510 (1995)

\bibitem{Malament2} D.~Malament, in {\em Handbook of the Philosophy of Physics, Eds.{} J.~Butterfield \& J.~Earman,\/}  229-273, Elsevier  (2006) 

\bibitem{Malament3} D.~Malament, {\em Topics in the Foundations of General Relativity and Newtonian Gravitational Theory\/}, University of Chicago Press (2012)

\bibitem{Tipler} F.J.~Tipler, Am.~J.~Phys.~{\bf64}:1311-1315 (1996)

\bibitem{Straumann} C.~R\"uede \& N.~Straumann, Helv.~Phys.~Acta {\bf 70}:318-335 (1997)


\bibitem{Norton} J.D.~Norton, Philosophy of Science {\bf 62}(4):511-522 (1995)


\bibitem{Layzer} D.~Layzer, Ast.~J.~{\bf59}:268-270 (1954)

\bibitem{Cartan} \'E.~Cartan, Annales Scientifique de l'\'Ecole Normale Sup\'erieure {\bf 40}:325-412 (1923)

\bibitem{Cartan2} \'E.~Cartan, Annales Scientifique de l'\'Ecole Normale Sup\'erieure {\bf 41}:1-25 (1924)

\bibitem{Friedrichs} K.~Friedrichs, Mathematische Annalen, {\bf98}:566-575 (1928)

\bibitem{Trautman} A.~Trautman, in {\em Lectures on General Relativity, Eds.{} S.~Deser \& K.W.~Ford,\/} Englewood Cliffs N.J., Prentice Hall, 1-248 (1965)

\bibitem{Dixon} W.G.~Dixon, Commun.~Math.~Phys.~{\bf 45}:167-182 (1975)

\bibitem{Iihoshi} M.~Iihoshi, S.V.~Ketov \& A.~Morishita,  Prog.~Theor.~Phys.~{\bf118}(3):475-489 (2007)

\bibitem{Ibison} M.~Ibison, 	J.~Math.~Phys, {\bf48}:122501 (2007) 



\end{thebibliography}
\end{document}